\documentclass[prl,twocolumn,showpacs,preprintnumbers,amsmath,amssymb,superscriptaddress]{revtex4-1}

\usepackage{color}
\usepackage{bm}                      
\usepackage{mathptmx}                
\usepackage{epsfig}

\usepackage{graphicx}
\usepackage{dcolumn}
\usepackage{microtype}
\usepackage[marginal]{footmisc}

\begin{document}

\preprint{APS/123-QED}

\title{Shape of a droplet on a surface in the presence of an external field and its critical disruption condition}

\author{Jing Li}%
\affiliation{Fujian Provincial Key Lab for Soft Functional Materials Research, Research Institute for Biomimetics and Soft Matter, Department of Physics, College of Physical Science and Technology, Xiamen University, Xiamen 361005, People's Republic of China}%
\author{Kaiqiang Wen}
\affiliation{Micro- and Nanotechnology Research Center, State Key Laboratory for Manufacturing Systems Engineering, Xi'an Jiaotong University, Xi'an, Shaanxi, 710049, China}
\author{Ke Xiao}
\affiliation{Fujian Provincial Key Lab for Soft Functional Materials Research, Research Institute for Biomimetics and Soft Matter, Department of Physics, College of Physical Science and Technology, Xiamen University, Xiamen 361005, People's Republic of China}
\author{Xiaoming Chen}%
\email{xiaomingchen@xjtu.edu.cn}
\affiliation{Micro- and Nanotechnology Research Center, State Key Laboratory for Manufacturing Systems Engineering, Xi'an Jiaotong University, Xi'an, Shaanxi, 710049, China}%
\author{Chen-Xu Wu}
\email{cxwu@xmu.edu.cn}
\affiliation{Fujian Provincial Key Lab for Soft Functional Materials Research, Research Institute for Biomimetics and Soft Matter, Department of Physics, College of Physical Science and Technology, Xiamen University, Xiamen 361005, People's Republic of China}%

\date{\today}

\begin{abstract}

Due to the potential application of regulating droplet shape by external fields in microfluidic technology and micro devices, it becomes increasingly important to understand the shape formation of a droplet in the presence of an electric field. How to understand and determine such a deformable boundary shape at equilibrium has been a long-term physical and mathematical challenge. Here, based on the theoretical model we propose, and combining the finite element method and the gradient descent algorithm, we successfully obtain the droplet shape by considering the contributions made by electrostatic energy, surface tension energy, and gravitational potential energy. We also carry out scaling analyses and obtain an empirical critical disruption condition with a universal scaling exponent 1/2 for the contact angle in terms of normalized volume. The master curve fits both the experimental and the numerical results very well.

\end{abstract}

\maketitle


\section{\label{sec:introduction}Introduction}

A droplet placed on a surface in the presence of an electric field will deform due to electrical stress at its surface. Such a phenomenon was first observed by William Gilbert, a personal physician of Queen Elizabeth, four centuries ago\cite{freudenthal1983theory}. If a sufficiently strong electric field is applied, the stability of the fluid interface between the droplet and the surrounding air will be disrupted\cite{zeleny1917instability}, causing the droplet to emit a charged micro droplet jet\cite{taylor1964disintegration}\cite{collins2008electrohydrodynamic}. There have been numerous studies on such a stability. Taylor determined the stability limit of a conductive free-floating droplet in a uniform electric field with experiments and size analysis with its shape assumed spheroidal\cite{wilson1925bursting}. The experiments carried out later on by Zeleny\cite{zeleny1917instability} and Taylor\cite{taylor1964disintegration,cloupeau1989electrostatic} demonstrated that, once the stability limit of the liquid droplet is exceeded, the unstable droplet becomes conical, now known as Taylor cones, with a stream or finite jet emitting from its tip. Such a spray phenomenon is the foundation of later high-resolution printing\cite{park2007high}, air purification\cite{jaworek2013submicron}, spray mass spectrometry technology\cite{fenn1989electrospray}, and ion beam generation technology\cite{krohn1975ion}. Besides, Electrospinning is also a related process in which a charged droplet emits fine liquid filaments from its tip under a high static voltage, which can be used to manufacture composite materials\cite{visser2015reinforcement}, nanogenerators\cite{wang2018polymer}, and drug carrier fibers\cite{kataria2014vivo}. In addition, droplet manipulation using electric field below the critical limit can be applied to control mixing and coalescence of emulsion droplets\cite{abate2010high}\cite{eow2002electrostatic}, as well as to drive droplet deformation, motion, merging, and splitting\cite{liu2023experimental}\cite{hartmann2022manipulation}, which has great significance in the manufacture of microfluidic devices.

Despite the great interest in stability of droplets, the analytical solution has not been derived for the general case, i.e. a droplet on a surface in the presence of a uniform external electric field, as shown in Fig.~\ref{schematic}. Mathematically  such a problem can be categorized as the second type of moving boundary problem (MBPs), with both its shape and distribution function undetermined\cite{vcanic2021moving}. The moving boundary here refers to the interface between two immiscible phases, i.e. the air and the liquid, which is hard to predict due to the competition between the interfacial interaction and the Coulomb interaction, which couple with each other in a very complicated way. Such an interplay will also be influenced by gravity if the droplet size is large enough in terms of capillary length. How to obtain the final equilibrium shape of a droplet in the presence of an electric field numerically and what determines the critical disruption voltage remains to be elucidated.

So far, the minimum energy principle is still the most powerful technique for solving equilibrium problems with changing boundaries as it transforms the complicated process of solving MBPs into an optimization of the total loss function based on a couple of restrictions and geometrical confinements, if necessary. For the theoretical solution of droplet shapes, some theoretical or numerical solutions can be obtained if a specific type of shapes is assumed, like the theoretical study of droplet evaporation\cite{man2016ring}\cite{man2017vapor} and droplet electrowetting\cite{tabassian2016graphene}. Such a treatment, however, is all based on the preassumed type of shapes, which is questionable for large droplet shape deformation due to the correctness of shape types selected. To address this, we expanded the droplet shape with azimuthal symmetry in terms of Legendre polynomials, and successfully derive a dynamic equation on shape deformation without any preassumptions\cite{xiao2022droplet}. Recently, a power law formula has been derived for the stability limit of a conducting droplet of small volume on a conducting surface, which enables us to predict the critical uniform electric field\cite{beroz2019stability}. Such a treatment ignores the contribution made by gravity, which holds only when the droplet size is much smaller than the capillary length.

In this article, we establish a theoretical model to describe the droplet shape and numerically solve the free energy by using droplet shape parameters. By applying the minimum energy principle, we obtain the equilibrium shapes of droplets under different external fields. The gradient descent Adam algorithm\cite{Kingma2014AdamAM} was introduced in the calculation so as to speed up the calculation. The present model also works in large volume droplet condition. The numerical results predicted by our model are compared with experimental ones. We expect this work can provide some helpful insights for the design of microfluidic devices when the voltage and other experimental parameters meet their requirements.

\section{Experiment}

\begin{figure}[htp]
  \includegraphics[width=\linewidth,keepaspectratio]{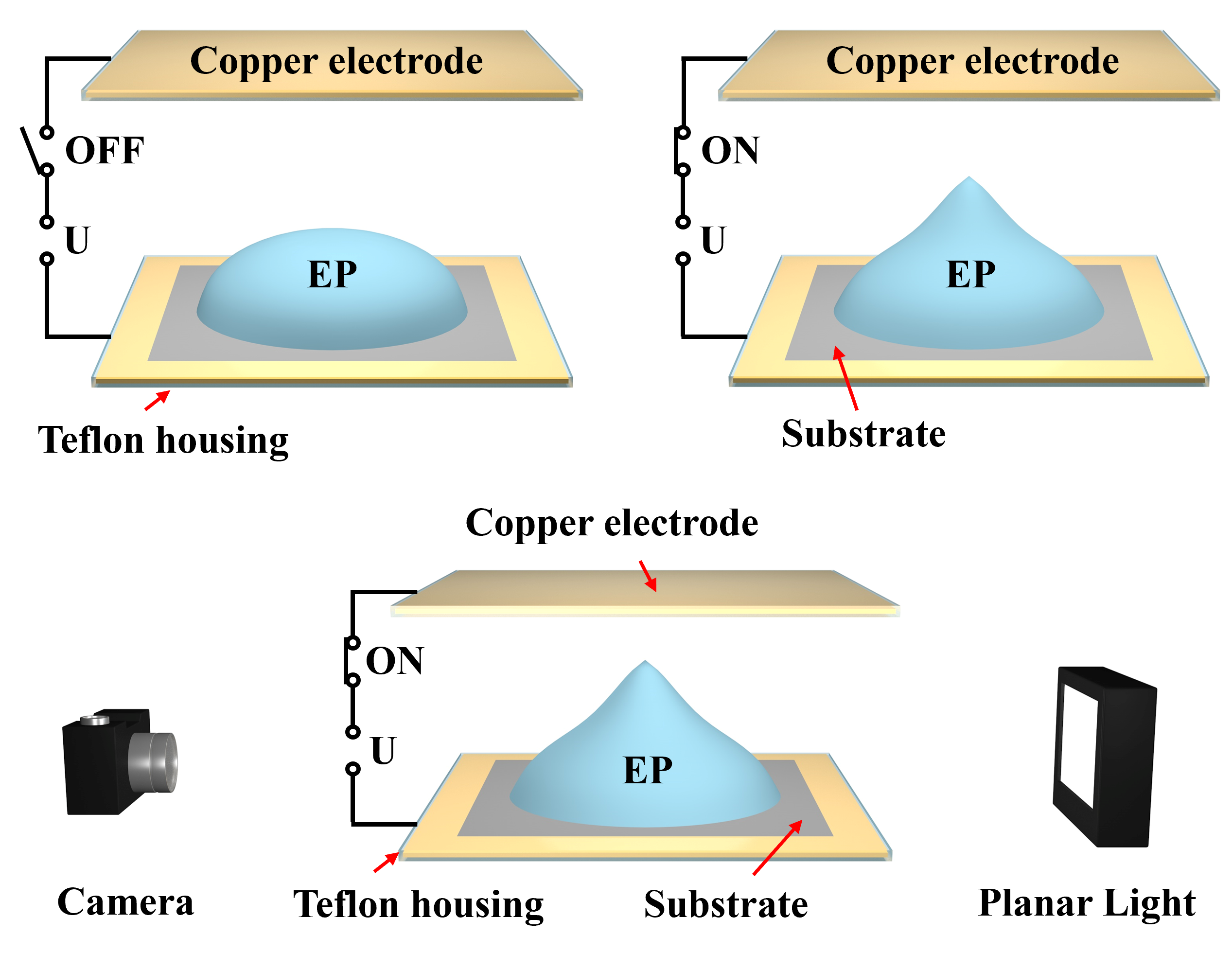}
  \caption{Schematic of a droplet placed on the lower plate of a capacitor in the presence of an external field.\label{schematic}}
\end{figure}

To carry out the experiments, we use a droplet of volume 15 $\mu$L of epoxy resin (EP; E51, 0.98 g/cm$^3$) supplied by Shanghai Maclean Biochemical Co. The process of droplet deformation under the application of an electric field can be summarized as follows: Initially, the substrate is fixed on the surface of the lower electrode. Then a droplet is deposited onto the substrate, and the upper electrode is directly snapped onto the surface of the lower electrode, maintaining a 5mm gap. Subsequently, signals from a function/arbitrary waveform generator (RIGOL, DG1022) are transmitted to a voltage amplifier/controller (TREK COR-A-TROL, model 610B). This controller amplifies and directs the voltage across the electrodes, establishing a spatial electric field, as shown in Fig.~\ref{schematic}. A dynamic contact angle meter (OTC50EC, Data Physics Instruments, Germany) is used to quantitatively assess the contact angle and the shape of the droplet. For comparative analysis, the wetting behavior of the droplet on the substrate was also observed in the absence of an electric field. After selecting the droplet material and substrate material, once an external electric field is applied, due to the combined effect of electric field polarization, gravitational effect and surface tension, the droplet will undergo various degrees of deformation. Due to the rapid response of droplets and the low ambient temperature during the experiment, the evaporation effect can be ignored.

\section{\label{sec:Theoretical model}Theoretical model}

The equilibrium state of the droplet as shown in Fig.~\ref{schematic} is reached via a competition between the electrostatic interaction, the interfacial tension and the gravitational potential. Theoretically, the total free energy of the system consists of three parts: the surface energy $F^{\rm S}$, the electrostatic energy $F^{\rm E}$, and the gravitational potential $F^{\rm G}$, i.e.,
\begin{equation}
F = F^{\rm S}+F^{\rm E}+F^{\rm G}.
\end{equation}
The surface energy existing at the three interfaces is given by
\begin{equation}
F^{\rm S} = \sigma_{\rm la}S_{\rm la}+\sigma_{\rm ls}S_{\rm ls}+\sigma_{\rm sa}S_{\rm sa},
\end{equation}
where $\sigma_{\rm la},\sigma_{\rm ls}$, and $\sigma_{\rm sa}$ represent the surface energies of the liquid-air, the liquid-substrate and the substrate-air, respectively, and $ S_{\rm la}, S_{\rm ls}$, and $S_{\rm sa}$ corresponds to their contact surface areas. Such a surface energy expression can be rewritten in terms of contact angle(CA) $\theta_{\rm e}$:
\begin{equation}
F^{\rm S}=\sigma_{\rm la}S_{\rm la}+\sigma_{\rm la}\cos\theta_{\rm e} S_{\rm ls}.
\end{equation}
Due to the large difference of dielectric constant, the electrostatic energy $F^{\rm E}$ is mainly contributed by the interaction between the polarization of liquid droplet and the external field, which is given by \cite{taylor1964disintegration,LANDAU198434,cheng1984deformation}
\begin{equation}
F^{\rm E}=-\frac{1}{2}\int_{\rm inside}E_0\cdot\left(\epsilon_0\epsilon_r^{\rm i}E_1-\epsilon_0\epsilon_{\rm r}^{\rm e}E_1\right)dV,
\end{equation}
where the integral is performed over the droplet volume. Here $\epsilon_0$ is the permittivity in a vacuum, $\epsilon_{\rm r}^{\rm i}$ and $\epsilon_{\rm r}^{\rm e}$ are the relative permittivity of liquid and air, respectively, $E_0$ is the uniform electric field when there is no droplet placed on the plate with voltage on, and $E_1$ is the electric field inside the droplet. On the other hand, it is easy to find out that the droplet size we consider is larger than the capillary length,
\begin{equation}
l_{\rm c}=\sqrt{\frac{\sigma_{\rm la}} {\rho_{\rm l} g}},\label{capillary length}
\end{equation}
the gravitational potential
\begin{equation}
F^{\rm G}=mgh_{\rm c},
\end{equation}
 needs to be taken into consideration. Here $m$ is the mass of droplet, $g$ is the gravitational acceleration, $\rho_{\rm l}$ is the density of liquid, and $h_{\rm c}$ is the height of the center of mass from the substrate.

Here it should be noted that it’s hard to obtain an analytical solution for the electrostatic energy as the boundary on which the calculation relies is deformable. The electrostatic energy comes from the interaction between the polarization of the droplet and the applied external electric field. Such an energy is regulated by deformable shape of the droplet, which is unknown initially and difficult to depict.

\begin{figure}[htp]
  \includegraphics[width=\linewidth,keepaspectratio]{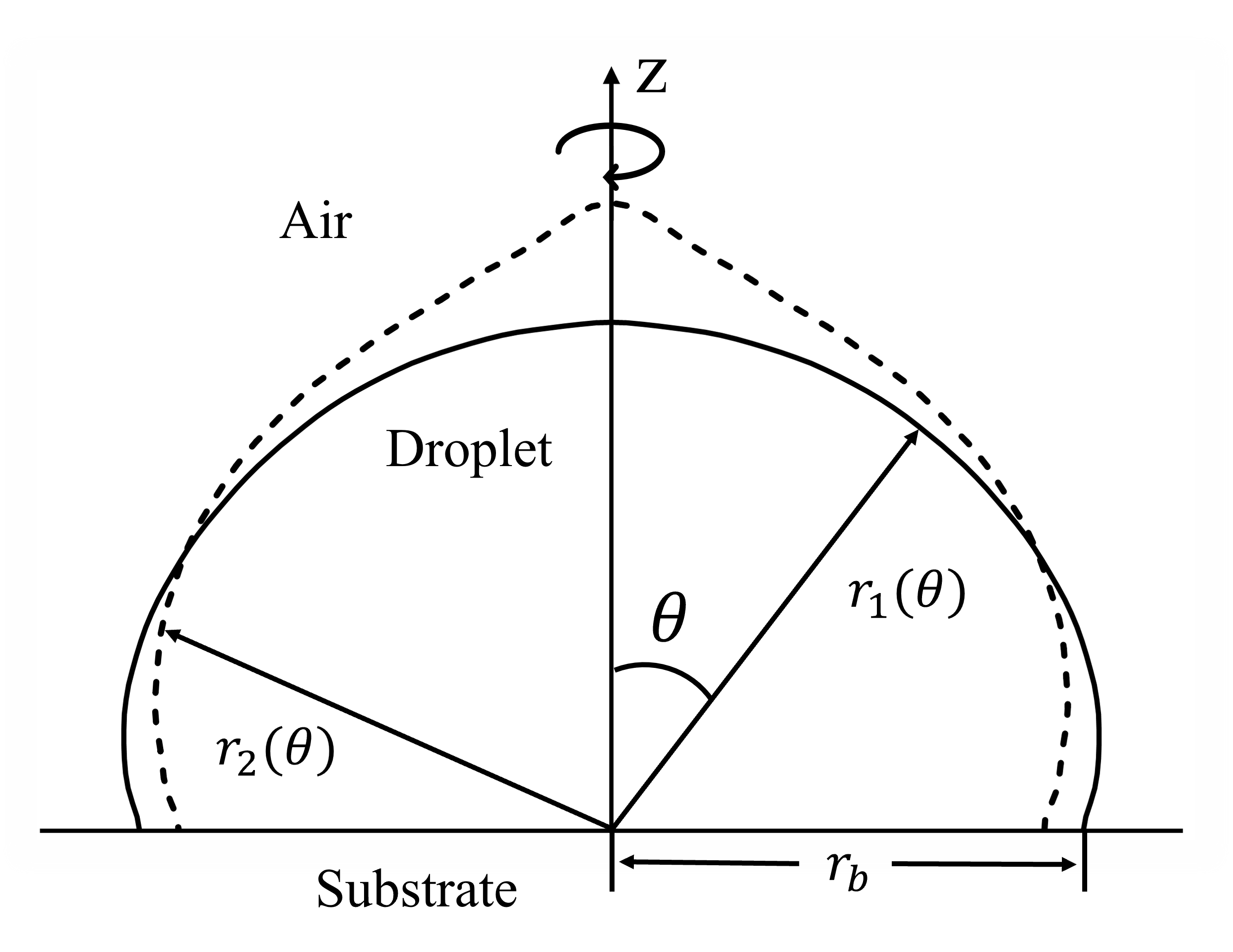}
  \caption{Meridian profile of a droplet with definitions of its parameters.}\label{meridian profile}
\end{figure}

Conventionally, to obtain the electrostatic energy of the system, we need to know its boundary shape, which is usually fixed. However, in this system, as the droplet shape is changeable due to its strong fluidity, a special approach is needed to characterize its shape. Because of the rotational symmetry and orthogonal properties of Legendre series, the shape of an axisymmetric drop can be represented by a weighted sum of the Legendre polynomials as \cite{oh2008shape,oh2010analysis}:
\begin{equation}
r = \sum_{n=0}^{\infty}a_nP_n\left(\cos\theta\right),\label{radial length}
\end{equation}
where $r$ denotes the radial length, $\theta$ is the angle made between radial direction and $z$ axis, $P_n$ is the $n$-th Legendre polynomial, and $a_n$ is the corresponding coefficient characterizing the contribution made by different profile mode of the droplet, as shown in Fig.~\ref{meridian profile}. Given the shape of the droplet in terms of the Legendre polynomial expansion coefficient $a_n$, the total energy can be written as a functional, depending on $a_n$.

\section{\label{sec:Numerical simulation}Numerical simulation}

\begin{figure}[htp]
  \includegraphics[width=0.8\linewidth,keepaspectratio]{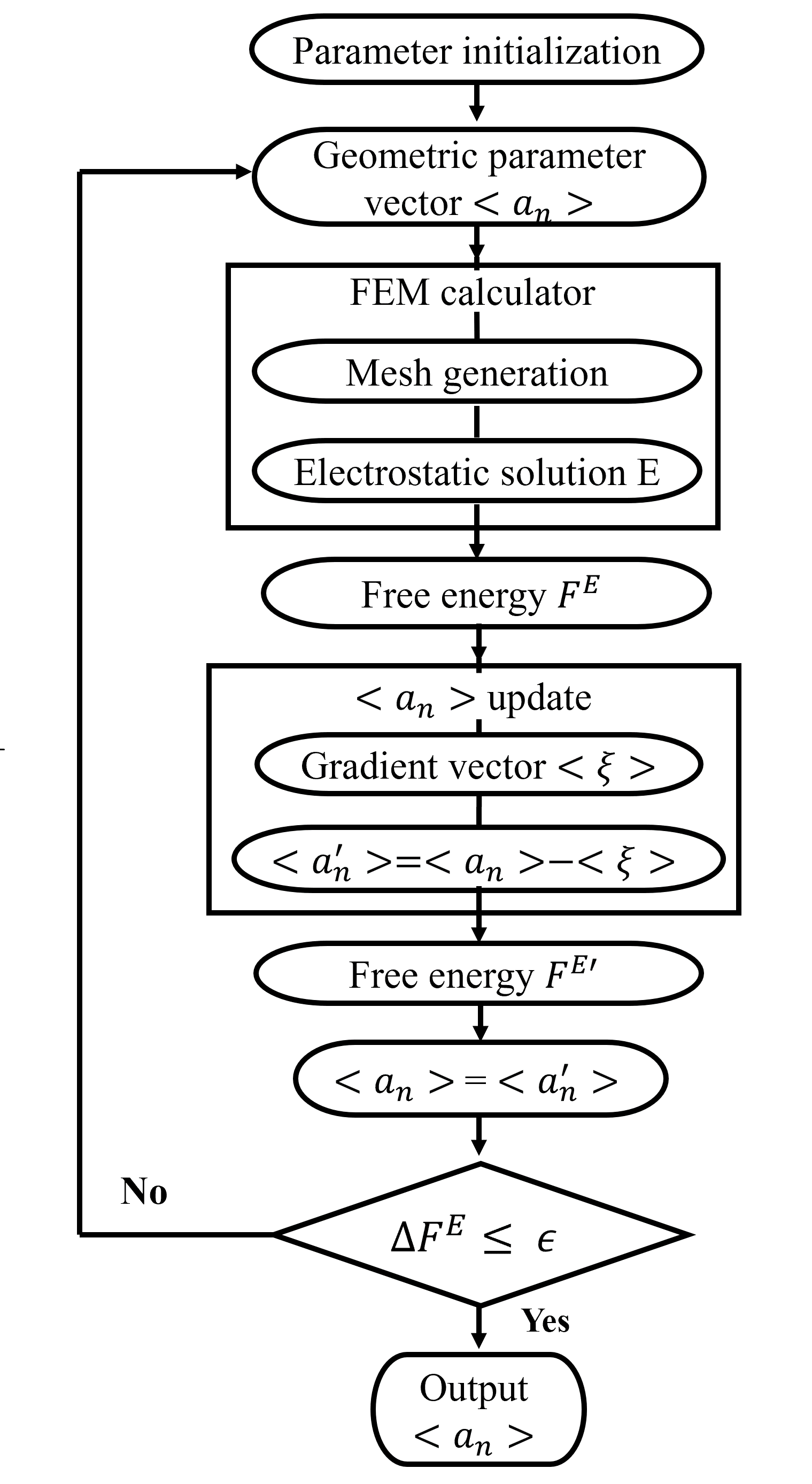}
  \caption{The simulation flow chart. }\label{flow chart}
\end{figure}

 The entire calculation process complies with the flow chart shown in Fig.~\ref{flow chart}.  Given the droplet material (surface tension $\sigma_{\rm la}$, the relative permittivity $\epsilon_{\rm r}^{\rm i}$), the droplet volume V, the applied electric field $E_0$ and the initial geometric shape vector $a_n$ describing the initial droplet shape, the total free energy of the system can be numerically solved by applying the finite element method. In order to optimize the free energy of the system, we start with an initial geometric shape vector, update it after calculating the total energy of the system. To get the fastest update rate, we use a gradient vector to update geometric shape vector so as to ensure a decreasing total energy at each loop. If the difference in free energy of before and after updating geometric shape vector is small enough, it is assumed that the system has converged to its equilibrium state with the droplet shape in equilibrium state defined by the output $a_n$. If not, one needs to continue the loop operation until the convergence is reached. To secure a solution that is physical, it is necessary to terminate the simulation so as to avoid the short circuit due to the contact of the jetting stream from the
 north pole of the droplet to the upper plate of the capacitor, if the voltage applied is large enough. Here, the specific operation of geometric shape vector updating is mainly referenced from Adam algorithm\cite{Kingma2014AdamAM}, a technique widely used in machine learning, and the finite element method is relied on COMSOL. Based on the algorithm, the stable shapes of droplets in equilibrium state under different initial conditions can be obtained. Since the bottom radius of the resin droplets is basically non-slip in experiment, so a geometric constraint with constant bottom radius is introduced to the numerical calculation.

\begin{figure*}[htp]
  \includegraphics[width=0.95\linewidth,keepaspectratio]{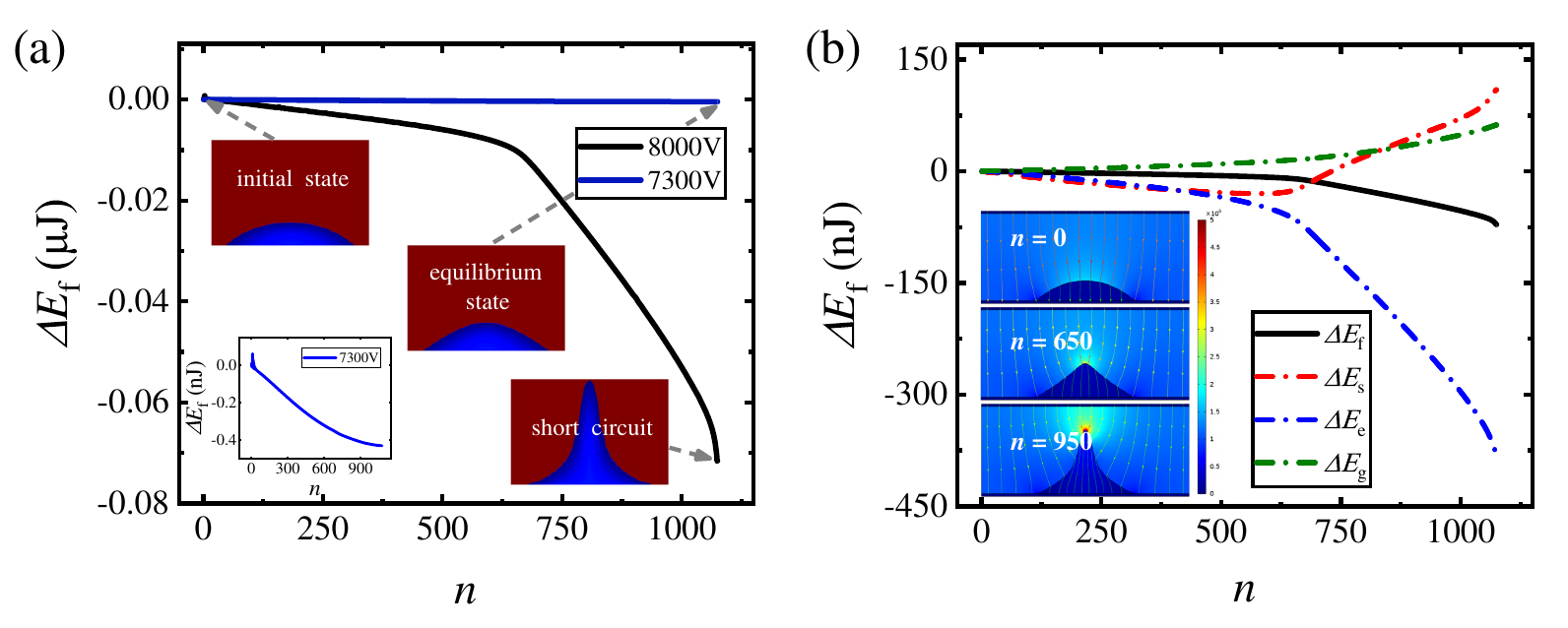}
  \caption{(a) Changes in free energy during the iteration process under the applied voltages of 7300V and 8000V with an initial CA equal to 40°. The blue area in the illustration is droplet and the red area is air. (b) Evolution of different energies during the iteration process with the applied voltage equal to 8000V. The illustration shows the spatial electric field distribution for different droplet deformations. }\label{calculation process}
\end{figure*}

In order to explore the behaviors of a droplet in the presence of an external field, a large number of numerical calculations and experiments were carried out for the comparisons between them. Figure~\ref{calculation process} shows how the total energy of the system converges when an external field is applied to a droplet with an initial shape of spherical cap.  In Fig.~\ref{calculation process}(a), when an external voltage of 7300V lower than the critical voltage is applied, the free energy of the system quickly converges to a stable value, with an energy change in a scale of nano joules, as shown by the blue curve in the inset at the bottom left corner. Here $n$ in the abscissa indicates the iteration number. The stable shape of droplet in equilibrium state is only slightly higher than the starting height of the north pole. When the applied voltage exceeds a certain threshold, such as 8000V, the contraction effect due to surface tension and gravitational force on the droplet cannot resist the polarization effect of the electric field on the droplet elongation, resulting in a continuously decreasing free energy curve (black line) until the system is short circuited. The jetting stream from the north pole of the droplet elongates upon iteration until it contacts the upper plate of the capacitor. As shown in the illustration in the bottom right corner of Fig.~\ref{calculation process}(a), where the blue droplet changes from its starting spherical crown shape to a tower-shaped during the iteration process, ending in a direct contact with the upper plate.

The impact on droplet shape of gravity effect has to be taken into consideration as the droplet volumes we study all exceed 15 $\mu L$, corresponding to  a droplet size of 1.93 mm, which is larger than the capillary length 1.75 mm estimated by Eq.~(\ref{capillary length}), given $\sigma_{\rm la}=30$$\rm mN/m$ and $\rho_{\rm l}=0.98$$\rm g/cm^3$. During the optimization process, the surface energy of the system tends to shrink the surface area of droplet, resulting in a decrease in system energy. At the same time, the electrostatic energy of the system tends to polarize the droplet, causing a local generation of large curvature in order to induce a local stronger electric field intensity distribution. Such a concentrated electric field built upon large curvature tends to reduce the electrostatic energy contribution to the system, which is contrary to the effect of surface energy. The gravitational force tries to lower the droplet as flat as possible, a tendency hindering the protrusion at the north pole of the droplet. When the applied bias voltage is lower than the critical voltage, a sufficiently large concentrated electric field cannot be built to overcome the contraction due to surface tension and flattening of droplet due to gravity field. A competition between them results in a compromise of a stable droplet shape at equilibrium. However, if the applied bias voltage exceeds the critical voltage, a spatially concentrated curvature will be generated at the top of the droplet during the optimization process of electrostatic energy, as shown in the illustration of $n=650$ in Fig.~\ref{calculation process}(b). At this moment, it is hard for the contraction of surface tension to hinder the stretching of the top of the droplet by the electrostatic force. The concentrated electric field and curvature at the top of the droplet will continue to be intensified, until a stream is jetted from this point to short circuit the system, as shown in the illustration for $n=950$. During this iteration process, the decrease in electrostatic energy of the system is sufficient to compensate for the increase of surface energy and gravitational potential energy, making the droplet approach to and reach the upper plate of the capacitor.

\section{Results and Discussions}

\begin{figure*}[htp]
  \centering
  \includegraphics[width=0.95\linewidth,keepaspectratio]{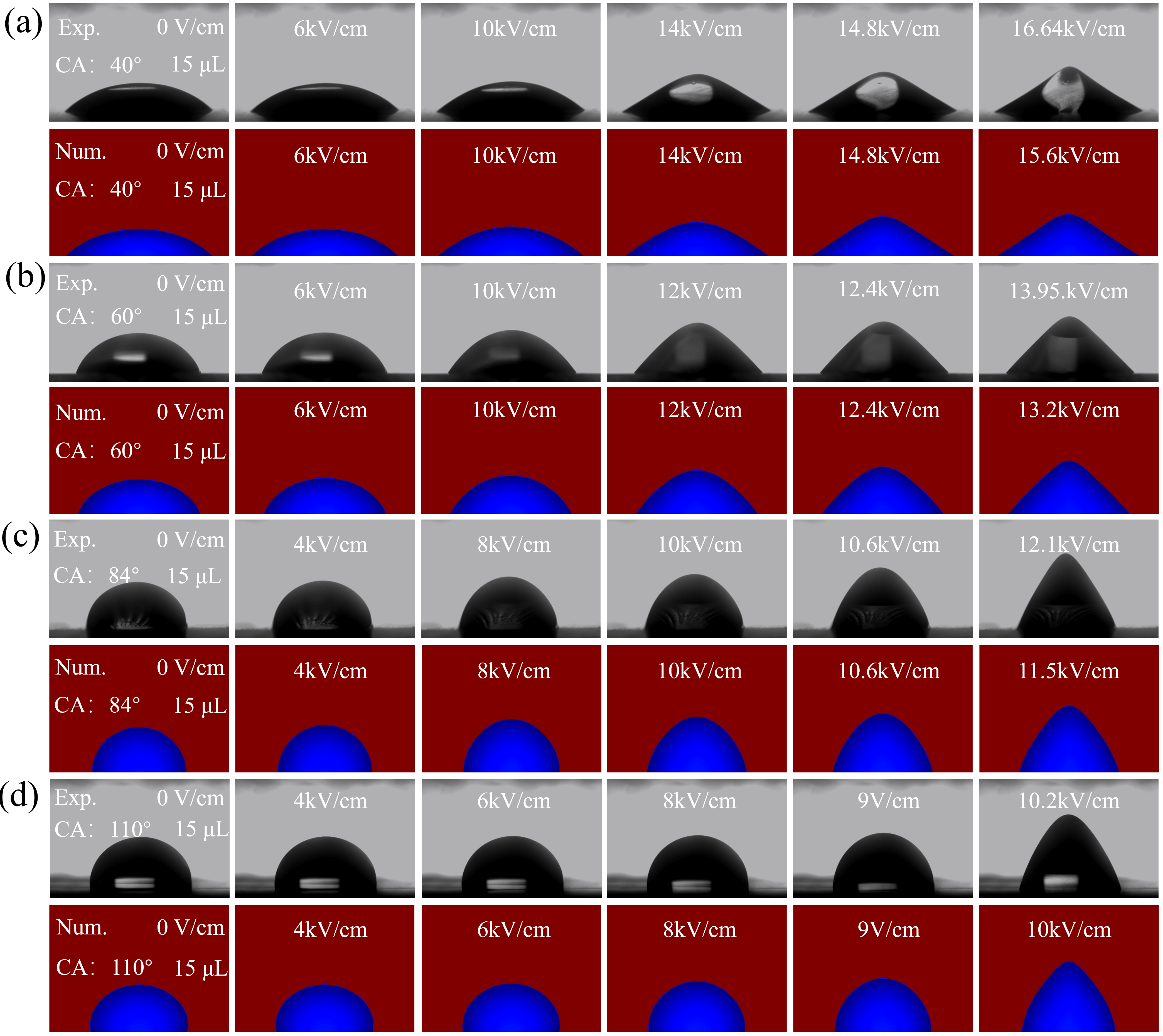}
  \caption{Meridian profiles of EP droplets with a limited droplet volume of 15 microliter in equilibrium state under initial contact angles (CAs) of (a) 40$^\circ$, (b) 60 $^\circ$, (c) 84$^\circ$, (d) 110$^\circ$, in the presence of various external electric fields, where the electric fields in the far-right figures correspond to the critical ones.}\label{comparison1}
\end{figure*}

Figure~\ref{comparison1} displays the changes in droplet shapes, with initial contact angles of 40$^\circ$, 60$^\circ$, 84$^\circ$, and 110$^\circ$, as they transition to equilibrium states under the action of a spatial electric field, comparing the numerical and experimental morphologies of the droplets. The droplet (EP) volume is maintained at 15 $\mu L$ throughout the experiment, with adjustments to the contact angle facilitated by surface treatments of the substrate. The comparison between experimental and computed results under identical conditions reveals a close similarity between the shapes of droplets obtained experimentally and those predicted numerically. Minor shape discrepancies may be attributed to the non-continuity of fluid in calculations and experimental errors. It is also found that as the applied electric field increases, the equilibrium droplet shape gradually transforms to a Taylor cone shape. The far-right figures in Fig.~\ref{comparison1} correspond to the cases of critical electric field application, beyond which a short circuited behavior of liquid droplets occur due to their contact to the upper plate of the capacitor. The critical field strengths measured by experiments and simulations are very close, indicating the feasibility and the accuracy of both the theoretical model and the algorithm we proposed.

\begin{figure}[htp]
  \includegraphics[width=\linewidth,keepaspectratio]{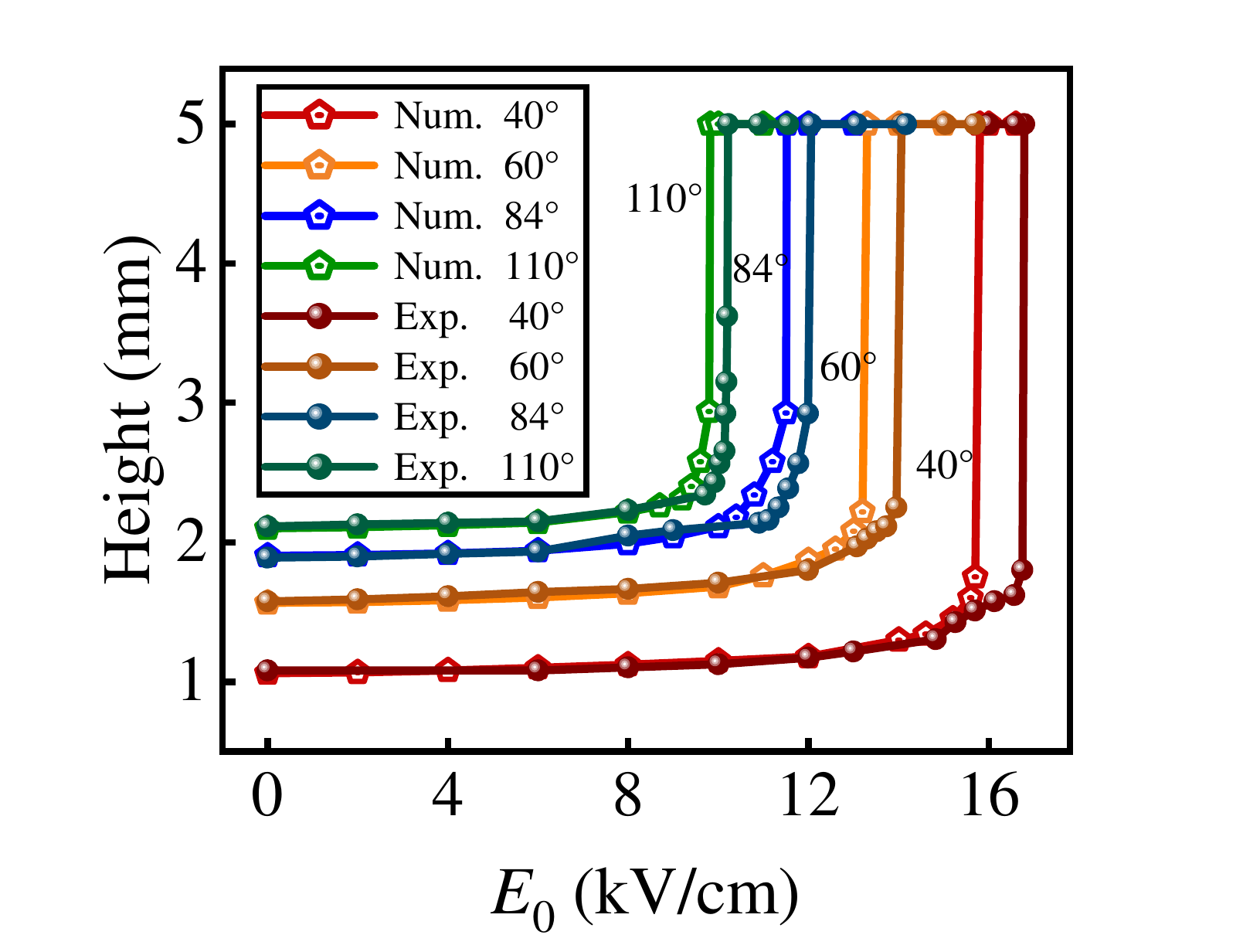}
  \caption{A comparison of EP droplet height under different contact angles in the presence of an external electric field between experiments and simulations.}\label{comparison2}
\end{figure}

In order to demonstrate the closeness of droplet shapes in equilibrium between experiments and calculations, we make a comparison of EP droplet height under different contact angles in the presence of an external field, as shown in Fig.~\ref{comparison2}, where a good agreement between simulations and experiments can be found. When the applied voltage is small, the polarization effect of the electric field impact on the droplets is weak, and one sees a slight increase of the height of the droplets. If the applied voltage is further increased, the energy contribution made by the electrostatic interaction increases, which requires more energy contributed by surface tension and gravity to balance against. As the droplet volume is fixed, the only tactic to realize this is to increase the local curvature, especially the height at the top of the droplet due to the symmetry of the system. Such a deformation can concentrate the electric field to the north pole of the droplet while keeping the total electrostatic energy of the system at the same level. When the applied voltage exceeds the critical field strength, the large electrostatic energy tends to accelerate such a tendency, even creating a stream of liquid jetting toward the upper plate of the capacitor. It is also shown in the figure that as the starting contact angle increases, the measured critical field strength gradually decreases. For the same droplet volume, small contact angle means large surface area and low center of mass, which therefore requires large critical electric field to break the energetic balance between them. The deformation of the droplet is a parameter to regulate the spatial distribution of the electrostatic energy, and the balance point between the two groups of energetic opponents as well.

\begin{figure}[htp]
  \includegraphics[width=\linewidth,keepaspectratio]{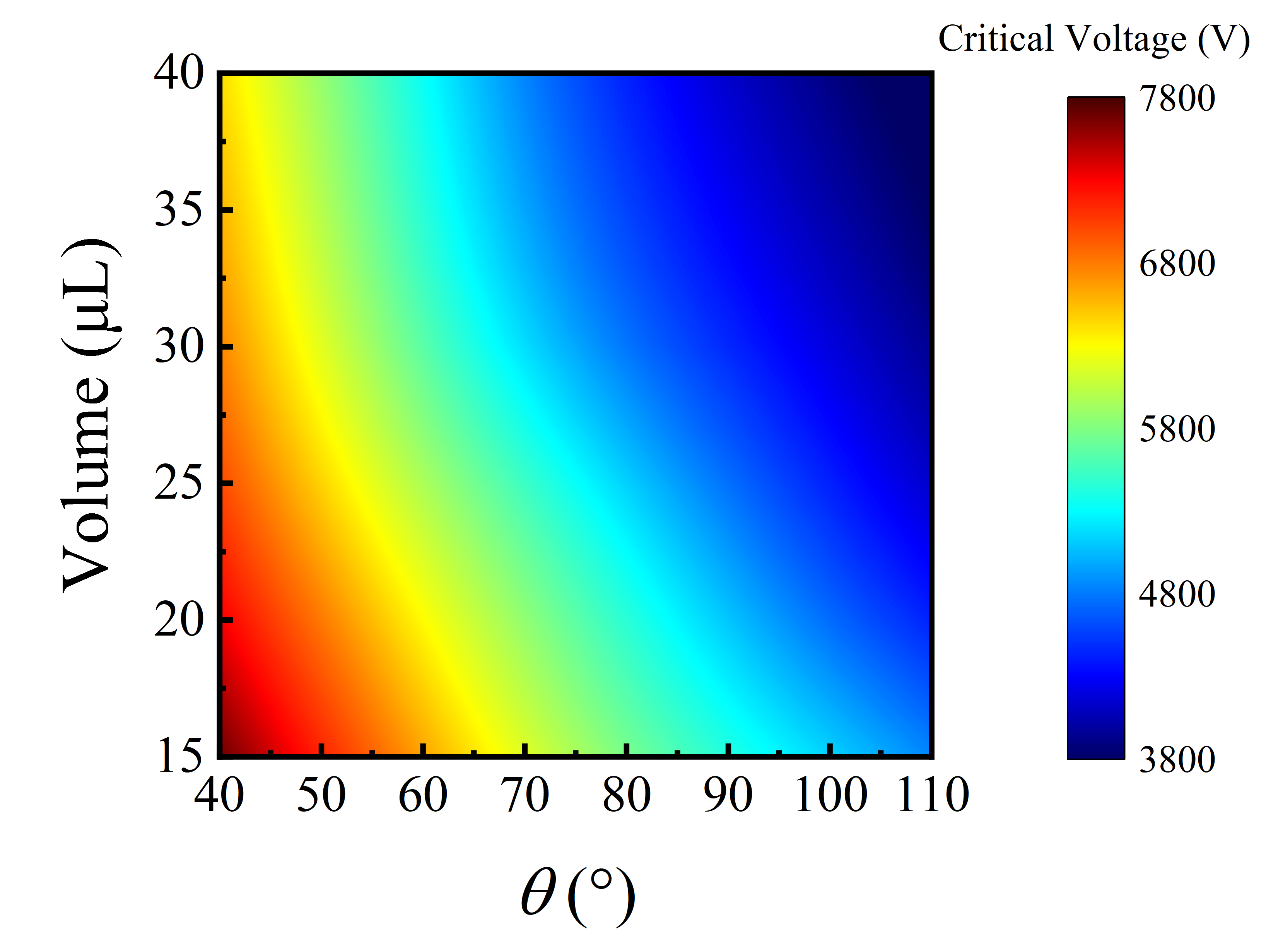}
  \caption{Dependence of critical voltage on droplet volume and contact angle $\theta$. The distance between the two parallel plates is $d=5$mm.}\label{Phase diagram}
\end{figure}

Via the experiments we carried out, we have already known that the critical voltage depends on both contact angle and volume, which are closely related to gravitational contribution. In order to systematically study the competitive relationship between surface tension energy, gravitational potential and electrostatic energy in critical state of such kind system, we  plot the critical voltage as a function of initial CA and droplet volume, as shown in Fig.~\ref{Phase diagram}(a). A substrate with weak hydrophilicity(large CA) requires a small critical field to short circuit the system. Even though a large droplet volume enhances both the gravity potential and the surface tension, it also shortens the effective distance between the conductive medium and the upper plate, and therefore reduces the threshold voltage to balance against the contributions made by gravity and surface tension. Such a correlation can help us choose a suitable substrate material and an appropriate droplet size for the lower plate of capacitors in the design of microfluidic instruments, so as to achieve a faster response under large applied voltages and avoid equipment short circuits.

\begin{figure}[htp]
  \centering
  \includegraphics[width=\linewidth,keepaspectratio]{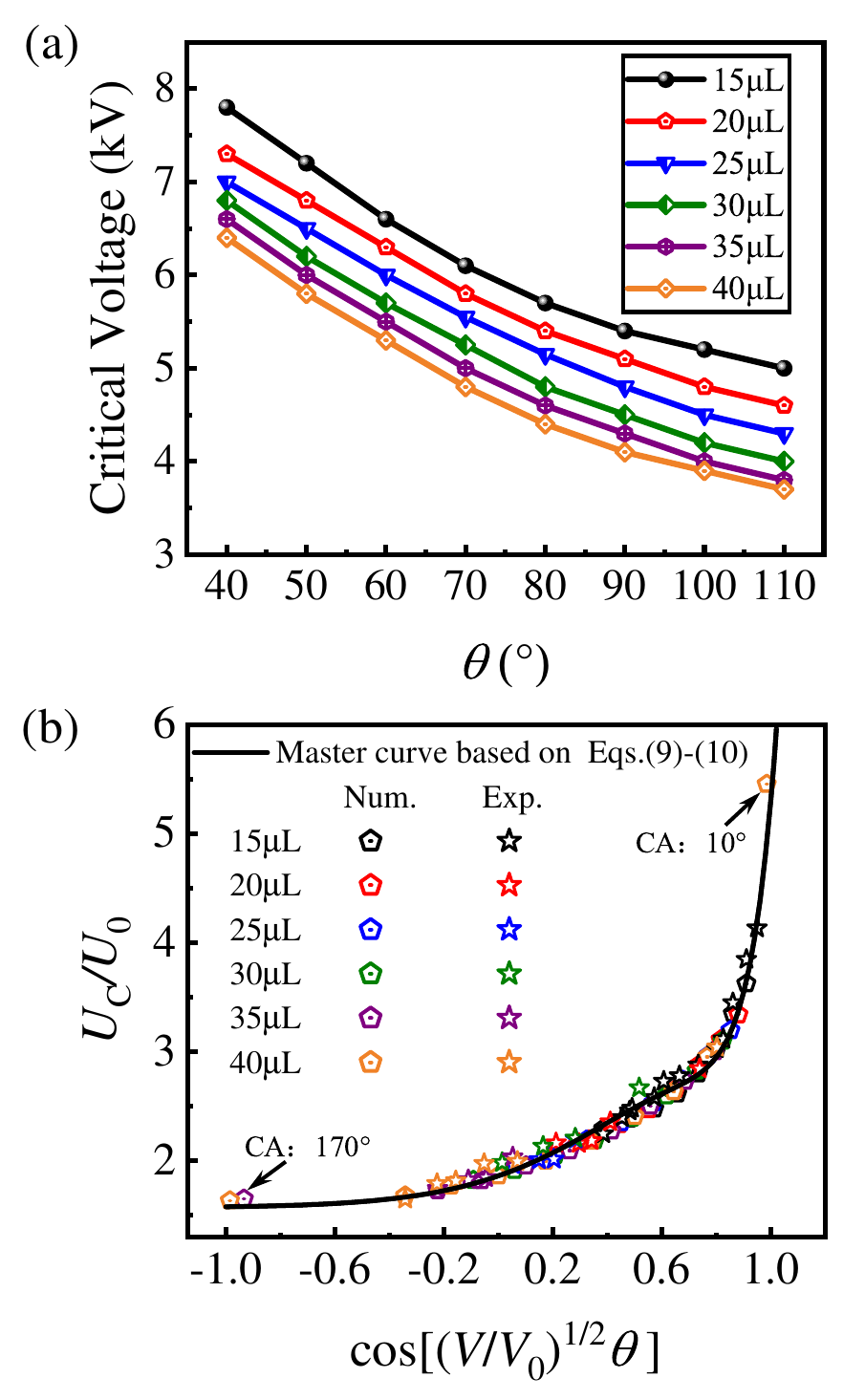}
  \caption{(a) Critical voltage as a function of contact angle $\theta$ based on numerical calculations and (b) its corresponding master curve. The distance between the two parallel plates is $d=5$mm.}\label{master curve}
\end{figure}

Even though Fig.~\ref{master curve}(a) depicts the critical disruption voltage as a function of contact angle for different droplet sizes, we aims to obtain a master curve for them. In order to achieve this, we have to find the intrinsic physical quantities hidden behind to rescale Fig.~\ref{master curve}(a). Here we use the typical length, i.e. the capillary length $l_{\rm c}$ given by Eq.~(\ref{capillary length}), to define a typical droplet surface area $l_{\rm c}^{\rm 2}$ and a typical droplet volume $l_{\rm c}^{\rm 3}$, respectively. Given these, a competition between the electrostatic energy and the surface tension leads to a typical intrinsic voltage:
\begin{equation}
U_0= d\sqrt{\frac{\sigma_{\rm la}}{\epsilon_0\epsilon_r^{\rm i} l_c}},\label{Intrinsic voltage}
\end{equation}
which is about 2200V based on the parameters given in system. Given $U_0$, the universal master curve for Fig.~\ref{master curve}(a) should be built among the dimensionless voltage $U_{\rm c}/U_0$, volume $V/V_0$ (for simplicity, we let $V_0=7.5l_{\rm c}^3=40 \mu L$ in this paper) and contact angle $\theta$, where $U_{\rm c}/U_0$ shows the competition between the Coulomb interaction and the surface tension, and $V/V_0$ and the contact angle overall reflect the contribution made by gravity. Rescaling  Fig.~\ref{master curve}(a) with normalized volume $V/V_0$ leads to a master curve, as shown in Fig.~\ref{master curve}(b), given by
\begin{equation}
\frac{U_c}{U_0} = f[\cos((V/V_0)^{1/2}\theta)],\label{critical condition}
\end{equation}
where a scaling exponent 1/2 is found for the contact angle $\theta$. More precisely, the function $ f(x)$ can be empirically written as a coupling between an exponential type and a polynomial one:
\begin{equation}
f(x) = \biggl(\frac{1}{3}x^2-\frac{3}{5}x
+0.293\biggr) \exp(5x) + 1.57.\label{f function}
\end{equation}
Equation~(\ref{critical condition}) together with Eq.~(\ref{f function}) gives the critical condition on the disruption of a droplet in the presence of an external electric field. As shown in Fig.~\ref{master curve}(b), the master curve based on Eqs.~(\ref{critical condition}) and (\ref{f function}) fits well with both numerical and experimental results, even in extreme cases of superhydrophilic substrates (CA: 10$^\circ$) and the superhydrophobic substrates (CA: 170$^\circ$). Based on the empirical formulae Eqs.~(\ref{critical condition}) and (\ref{f function}), researchers can predict the exact electric field strength at which droplets disrupt or below which maintain their stability. Such empirical expressions work for both cases: droplets placed on solid substrates and floating droplets exposed to an electric field (CA: 90$^\circ$).

\section{Conclusion}

In summary, based on the proposed theoretical model, we successfully obtain the shape of a droplet on a surface in the presence of an external field, by combining the finite element method and the gradient descent algorithm to solve such a second-type moving boundary problem. A very good agreement between experiments and numerical simulations has been found, indicating the validity of the present model. Moreover, an empirical expression on how the critical voltage depends on the droplet volume and the contact angle has been obtained. As indicated by the simulations, selecting materials with large hydrophobicity and increasing the volume of regulated droplets can significantly reduce the critical voltage of short circuit, providing a practical design criterion in the applications such as the industrial-scale electrospinning, electrostatic filtration, demulsification etc.

\section{Acknowledgments}
The theoretical model was constructed with its analyses and numerical calculations performed by Chen-Xu Wu's group and the experimental data were provided by Xiaoming Chen's group. Jing Li and Kaiqiang Wen are treated as co-first authors. We acknowledge financial support from National Natural Science Foundation of China under Grant No.12174323, 52175544 and 111 project B16029.


\begin{thebibliography}{28}%
\makeatletter
\providecommand \@ifxundefined [1]{%
 \@ifx{#1\undefined}
}%
\providecommand \@ifnum [1]{%
 \ifnum #1\expandafter \@firstoftwo
 \else \expandafter \@secondoftwo
 \fi
}%
\providecommand \@ifx [1]{%
 \ifx #1\expandafter \@firstoftwo
 \else \expandafter \@secondoftwo
 \fi
}%
\providecommand \natexlab [1]{#1}%
\providecommand \enquote  [1]{``#1''}%
\providecommand \bibnamefont  [1]{#1}%
\providecommand \bibfnamefont [1]{#1}%
\providecommand \citenamefont [1]{#1}%
\providecommand \href@noop [0]{\@secondoftwo}%
\providecommand \href [0]{\begingroup \@sanitize@url \@href}%
\providecommand \@href[1]{\@@startlink{#1}\@@href}%
\providecommand \@@href[1]{\endgroup#1\@@endlink}%
\providecommand \@sanitize@url [0]{\catcode `\\12\catcode `\$12\catcode `\&12\catcode `\#12\catcode `\^12\catcode `\_12\catcode `\%12\relax}%
\providecommand \@@startlink[1]{}%
\providecommand \@@endlink[0]{}%
\providecommand \url  [0]{\begingroup\@sanitize@url \@url }%
\providecommand \@url [1]{\endgroup\@href {#1}{\urlprefix }}%
\providecommand \urlprefix  [0]{URL }%
\providecommand \Eprint [0]{\href }%
\providecommand \doibase [0]{https://doi.org/}%
\providecommand \selectlanguage [0]{\@gobble}%
\providecommand \bibinfo  [0]{\@secondoftwo}%
\providecommand \bibfield  [0]{\@secondoftwo}%
\providecommand \translation [1]{[#1]}%
\providecommand \BibitemOpen [0]{}%
\providecommand \bibitemStop [0]{}%
\providecommand \bibitemNoStop [0]{.\EOS\space}%
\providecommand \EOS [0]{\spacefactor3000\relax}%
\providecommand \BibitemShut  [1]{\csname bibitem#1\endcsname}%
\let\auto@bib@innerbib\@empty
\bibitem [{\citenamefont {Freudenthal}(1983)}]{freudenthal1983theory}%
  \BibitemOpen
  \bibfield  {author} {\bibinfo {author} {\bibfnamefont {G.}~\bibnamefont {Freudenthal}},\ }\href@noop {} {\bibfield  {journal} {\bibinfo  {journal} {Isis}\ }\textbf {\bibinfo {volume} {74}},\ \bibinfo {pages} {22} (\bibinfo {year} {1983})}\BibitemShut {NoStop}%
\bibitem [{\citenamefont {Zeleny}(1917)}]{zeleny1917instability}%
  \BibitemOpen
  \bibfield  {author} {\bibinfo {author} {\bibfnamefont {J.}~\bibnamefont {Zeleny}},\ }\href@noop {} {\bibfield  {journal} {\bibinfo  {journal} {Phys. Rev.}\ }\textbf {\bibinfo {volume} {10}},\ \bibinfo {pages} {1} (\bibinfo {year} {1917})}\BibitemShut {NoStop}%
\bibitem [{\citenamefont {Taylor}(1964)}]{taylor1964disintegration}%
  \BibitemOpen
  \bibfield  {author} {\bibinfo {author} {\bibfnamefont {G.~I.}\ \bibnamefont {Taylor}},\ }\href@noop {} {\bibfield  {journal} {\bibinfo  {journal} {P Roy Soc A-math Phy}\ }\textbf {\bibinfo {volume} {280}},\ \bibinfo {pages} {383} (\bibinfo {year} {1964})}\BibitemShut {NoStop}%
\bibitem [{\citenamefont {Collins}\ \emph {et~al.}(2008)\citenamefont {Collins}, \citenamefont {Jones}, \citenamefont {Harris},\ and\ \citenamefont {Basaran}}]{collins2008electrohydrodynamic}%
  \BibitemOpen
  \bibfield  {author} {\bibinfo {author} {\bibfnamefont {R.~T.}\ \bibnamefont {Collins}}, \bibinfo {author} {\bibfnamefont {J.~J.}\ \bibnamefont {Jones}}, \bibinfo {author} {\bibfnamefont {M.~T.}\ \bibnamefont {Harris}},\ and\ \bibinfo {author} {\bibfnamefont {O.~A.}\ \bibnamefont {Basaran}},\ }\href@noop {} {\bibfield  {journal} {\bibinfo  {journal} {Nat. Phys.}\ }\textbf {\bibinfo {volume} {4}},\ \bibinfo {pages} {149} (\bibinfo {year} {2008})}\BibitemShut {NoStop}%
\bibitem [{\citenamefont {Wilson}\ and\ \citenamefont {Taylor}(1925)}]{wilson1925bursting}%
  \BibitemOpen
  \bibfield  {author} {\bibinfo {author} {\bibfnamefont {C.}~\bibnamefont {Wilson}}\ and\ \bibinfo {author} {\bibfnamefont {G.}~\bibnamefont {Taylor}},\ }in\ \href@noop {} {\emph {\bibinfo {booktitle} {Mathematical proceedings of the Cambridge philosophical society}}},\ Vol.~\bibinfo {volume} {22}\ (\bibinfo {organization} {Cambridge University Press},\ \bibinfo {year} {1925})\ pp.\ \bibinfo {pages} {728--730}\BibitemShut {NoStop}%
\bibitem [{\citenamefont {Cloupeau}\ and\ \citenamefont {Prunet-Foch}(1989)}]{cloupeau1989electrostatic}%
  \BibitemOpen
  \bibfield  {author} {\bibinfo {author} {\bibfnamefont {M.}~\bibnamefont {Cloupeau}}\ and\ \bibinfo {author} {\bibfnamefont {B.}~\bibnamefont {Prunet-Foch}},\ }\href@noop {} {\bibfield  {journal} {\bibinfo  {journal} {J Electrostat}\ }\textbf {\bibinfo {volume} {22}},\ \bibinfo {pages} {135} (\bibinfo {year} {1989})}\BibitemShut {NoStop}%
\bibitem [{\citenamefont {Park}\ \emph {et~al.}(2007)\citenamefont {Park}, \citenamefont {Hardy}, \citenamefont {Kang}, \citenamefont {Barton}, \citenamefont {Adair}, \citenamefont {Mukhopadhyay}, \citenamefont {Lee}, \citenamefont {Strano}, \citenamefont {Alleyne}, \citenamefont {Georgiadis} \emph {et~al.}}]{park2007high}%
  \BibitemOpen
  \bibfield  {author} {\bibinfo {author} {\bibfnamefont {J.-U.}\ \bibnamefont {Park}}, \bibinfo {author} {\bibfnamefont {M.}~\bibnamefont {Hardy}}, \bibinfo {author} {\bibfnamefont {S.~J.}\ \bibnamefont {Kang}}, \bibinfo {author} {\bibfnamefont {K.}~\bibnamefont {Barton}}, \bibinfo {author} {\bibfnamefont {K.}~\bibnamefont {Adair}}, \bibinfo {author} {\bibfnamefont {D.~K.}\ \bibnamefont {Mukhopadhyay}}, \bibinfo {author} {\bibfnamefont {C.~Y.}\ \bibnamefont {Lee}}, \bibinfo {author} {\bibfnamefont {M.~S.}\ \bibnamefont {Strano}}, \bibinfo {author} {\bibfnamefont {A.~G.}\ \bibnamefont {Alleyne}}, \bibinfo {author} {\bibfnamefont {J.~G.}\ \bibnamefont {Georgiadis}}, \emph {et~al.},\ }\href@noop {} {\bibfield  {journal} {\bibinfo  {journal} {Nat. Mater.}\ }\textbf {\bibinfo {volume} {6}},\ \bibinfo {pages} {782} (\bibinfo {year} {2007})}\BibitemShut {NoStop}%
\bibitem [{\citenamefont {Jaworek}\ \emph {et~al.}(2013)\citenamefont {Jaworek}, \citenamefont {Krupa}, \citenamefont {Sobczyk}, \citenamefont {Marchewicz}, \citenamefont {Szudyga}, \citenamefont {Antes}, \citenamefont {Balachandran}, \citenamefont {Di~Natale},\ and\ \citenamefont {Carotenuto}}]{jaworek2013submicron}%
  \BibitemOpen
  \bibfield  {author} {\bibinfo {author} {\bibfnamefont {A.}~\bibnamefont {Jaworek}}, \bibinfo {author} {\bibfnamefont {A.}~\bibnamefont {Krupa}}, \bibinfo {author} {\bibfnamefont {A.~T.}\ \bibnamefont {Sobczyk}}, \bibinfo {author} {\bibfnamefont {A.}~\bibnamefont {Marchewicz}}, \bibinfo {author} {\bibfnamefont {M.}~\bibnamefont {Szudyga}}, \bibinfo {author} {\bibfnamefont {T.}~\bibnamefont {Antes}}, \bibinfo {author} {\bibfnamefont {W.}~\bibnamefont {Balachandran}}, \bibinfo {author} {\bibfnamefont {F.}~\bibnamefont {Di~Natale}},\ and\ \bibinfo {author} {\bibfnamefont {C.}~\bibnamefont {Carotenuto}},\ }\href@noop {} {\bibfield  {journal} {\bibinfo  {journal} {J Electrostat}\ }\textbf {\bibinfo {volume} {71}},\ \bibinfo {pages} {345} (\bibinfo {year} {2013})}\BibitemShut {NoStop}%
\bibitem [{\citenamefont {Fenn}\ \emph {et~al.}(1989)\citenamefont {Fenn}, \citenamefont {Mann}, \citenamefont {Meng}, \citenamefont {Wong},\ and\ \citenamefont {Whitehouse}}]{fenn1989electrospray}%
  \BibitemOpen
  \bibfield  {author} {\bibinfo {author} {\bibfnamefont {J.~B.}\ \bibnamefont {Fenn}}, \bibinfo {author} {\bibfnamefont {M.}~\bibnamefont {Mann}}, \bibinfo {author} {\bibfnamefont {C.~K.}\ \bibnamefont {Meng}}, \bibinfo {author} {\bibfnamefont {S.~F.}\ \bibnamefont {Wong}},\ and\ \bibinfo {author} {\bibfnamefont {C.~M.}\ \bibnamefont {Whitehouse}},\ }\href@noop {} {\bibfield  {journal} {\bibinfo  {journal} {Science}\ }\textbf {\bibinfo {volume} {246}},\ \bibinfo {pages} {64} (\bibinfo {year} {1989})}\BibitemShut {NoStop}%
\bibitem [{\citenamefont {Krohn}\ and\ \citenamefont {Ringo}(1975)}]{krohn1975ion}%
  \BibitemOpen
  \bibfield  {author} {\bibinfo {author} {\bibfnamefont {V.~E.}\ \bibnamefont {Krohn}}\ and\ \bibinfo {author} {\bibfnamefont {G.~R.}\ \bibnamefont {Ringo}},\ }\href@noop {} {\bibfield  {journal} {\bibinfo  {journal} {Appl. Phys. Lett.}\ }\textbf {\bibinfo {volume} {27}},\ \bibinfo {pages} {479} (\bibinfo {year} {1975})}\BibitemShut {NoStop}%
\bibitem [{\citenamefont {Visser}\ \emph {et~al.}(2015)\citenamefont {Visser}, \citenamefont {Melchels}, \citenamefont {Jeon}, \citenamefont {Van~Bussel}, \citenamefont {Kimpton}, \citenamefont {Byrne}, \citenamefont {Dhert}, \citenamefont {Dalton}, \citenamefont {Hutmacher},\ and\ \citenamefont {Malda}}]{visser2015reinforcement}%
  \BibitemOpen
  \bibfield  {author} {\bibinfo {author} {\bibfnamefont {J.}~\bibnamefont {Visser}}, \bibinfo {author} {\bibfnamefont {F.~P.}\ \bibnamefont {Melchels}}, \bibinfo {author} {\bibfnamefont {J.~E.}\ \bibnamefont {Jeon}}, \bibinfo {author} {\bibfnamefont {E.~M.}\ \bibnamefont {Van~Bussel}}, \bibinfo {author} {\bibfnamefont {L.~S.}\ \bibnamefont {Kimpton}}, \bibinfo {author} {\bibfnamefont {H.~M.}\ \bibnamefont {Byrne}}, \bibinfo {author} {\bibfnamefont {W.~J.}\ \bibnamefont {Dhert}}, \bibinfo {author} {\bibfnamefont {P.~D.}\ \bibnamefont {Dalton}}, \bibinfo {author} {\bibfnamefont {D.~W.}\ \bibnamefont {Hutmacher}},\ and\ \bibinfo {author} {\bibfnamefont {J.}~\bibnamefont {Malda}},\ }\href@noop {} {\bibfield  {journal} {\bibinfo  {journal} {Nat. Commun.}\ }\textbf {\bibinfo {volume} {6}},\ \bibinfo {pages} {1} (\bibinfo {year} {2015})}\BibitemShut {NoStop}%
\bibitem [{\citenamefont {Wang}\ \emph {et~al.}(2018)\citenamefont {Wang}, \citenamefont {Wu}, \citenamefont {Pisignano}, \citenamefont {Wang},\ and\ \citenamefont {Persano}}]{wang2018polymer}%
  \BibitemOpen
  \bibfield  {author} {\bibinfo {author} {\bibfnamefont {A.~C.}\ \bibnamefont {Wang}}, \bibinfo {author} {\bibfnamefont {C.}~\bibnamefont {Wu}}, \bibinfo {author} {\bibfnamefont {D.}~\bibnamefont {Pisignano}}, \bibinfo {author} {\bibfnamefont {Z.~L.}\ \bibnamefont {Wang}},\ and\ \bibinfo {author} {\bibfnamefont {L.}~\bibnamefont {Persano}},\ }\href@noop {} {\bibfield  {journal} {\bibinfo  {journal} {J. Appl. Polym. Sci.}\ }\textbf {\bibinfo {volume} {135}},\ \bibinfo {pages} {45674} (\bibinfo {year} {2018})}\BibitemShut {NoStop}%
\bibitem [{\citenamefont {Kataria}\ \emph {et~al.}(2014)\citenamefont {Kataria}, \citenamefont {Gupta}, \citenamefont {Rath}, \citenamefont {Mathur},\ and\ \citenamefont {Dhakate}}]{kataria2014vivo}%
  \BibitemOpen
  \bibfield  {author} {\bibinfo {author} {\bibfnamefont {K.}~\bibnamefont {Kataria}}, \bibinfo {author} {\bibfnamefont {A.}~\bibnamefont {Gupta}}, \bibinfo {author} {\bibfnamefont {G.}~\bibnamefont {Rath}}, \bibinfo {author} {\bibfnamefont {R.}~\bibnamefont {Mathur}},\ and\ \bibinfo {author} {\bibfnamefont {S.}~\bibnamefont {Dhakate}},\ }\href@noop {} {\bibfield  {journal} {\bibinfo  {journal} {Int. J. Pharm.}\ }\textbf {\bibinfo {volume} {469}},\ \bibinfo {pages} {102} (\bibinfo {year} {2014})}\BibitemShut {NoStop}%
\bibitem [{\citenamefont {Abate}\ \emph {et~al.}(2010)\citenamefont {Abate}, \citenamefont {Hung}, \citenamefont {Mary}, \citenamefont {Agresti},\ and\ \citenamefont {Weitz}}]{abate2010high}%
  \BibitemOpen
  \bibfield  {author} {\bibinfo {author} {\bibfnamefont {A.~R.}\ \bibnamefont {Abate}}, \bibinfo {author} {\bibfnamefont {T.}~\bibnamefont {Hung}}, \bibinfo {author} {\bibfnamefont {P.}~\bibnamefont {Mary}}, \bibinfo {author} {\bibfnamefont {J.~J.}\ \bibnamefont {Agresti}},\ and\ \bibinfo {author} {\bibfnamefont {D.~A.}\ \bibnamefont {Weitz}},\ }\href@noop {} {\bibfield  {journal} {\bibinfo  {journal} {Proc. Natl. Acad. Sci. U.S.A.}\ }\textbf {\bibinfo {volume} {107}},\ \bibinfo {pages} {19163} (\bibinfo {year} {2010})}\BibitemShut {NoStop}%
\bibitem [{\citenamefont {Eow}\ and\ \citenamefont {Ghadiri}(2002)}]{eow2002electrostatic}%
  \BibitemOpen
  \bibfield  {author} {\bibinfo {author} {\bibfnamefont {J.~S.}\ \bibnamefont {Eow}}\ and\ \bibinfo {author} {\bibfnamefont {M.}~\bibnamefont {Ghadiri}},\ }\href@noop {} {\bibfield  {journal} {\bibinfo  {journal} {Chem. Eng. J.}\ }\textbf {\bibinfo {volume} {85}},\ \bibinfo {pages} {357} (\bibinfo {year} {2002})}\BibitemShut {NoStop}%
\bibitem [{\citenamefont {Liu}\ \emph {et~al.}(2023)\citenamefont {Liu}, \citenamefont {Hao}, \citenamefont {Li},\ and\ \citenamefont {Chen}}]{liu2023experimental}%
  \BibitemOpen
  \bibfield  {author} {\bibinfo {author} {\bibfnamefont {X.}~\bibnamefont {Liu}}, \bibinfo {author} {\bibfnamefont {G.}~\bibnamefont {Hao}}, \bibinfo {author} {\bibfnamefont {B.}~\bibnamefont {Li}},\ and\ \bibinfo {author} {\bibfnamefont {Y.}~\bibnamefont {Chen}},\ }\href@noop {} {\bibfield  {journal} {\bibinfo  {journal} {Fundamental Research}\ }\textbf {\bibinfo {volume} {3}},\ \bibinfo {pages} {274} (\bibinfo {year} {2023})}\BibitemShut {NoStop}%
\bibitem [{\citenamefont {Hartmann}\ \emph {et~al.}(2022)\citenamefont {Hartmann}, \citenamefont {Sch{\"u}r},\ and\ \citenamefont {Hardt}}]{hartmann2022manipulation}%
  \BibitemOpen
  \bibfield  {author} {\bibinfo {author} {\bibfnamefont {J.}~\bibnamefont {Hartmann}}, \bibinfo {author} {\bibfnamefont {M.~T.}\ \bibnamefont {Sch{\"u}r}},\ and\ \bibinfo {author} {\bibfnamefont {S.}~\bibnamefont {Hardt}},\ }\href@noop {} {\bibfield  {journal} {\bibinfo  {journal} {Nat. Commun.}\ }\textbf {\bibinfo {volume} {13}},\ \bibinfo {pages} {289} (\bibinfo {year} {2022})}\BibitemShut {NoStop}%
\bibitem [{\citenamefont {{\v{C}}ani{\'c}}(2021)}]{vcanic2021moving}%
  \BibitemOpen
  \bibfield  {author} {\bibinfo {author} {\bibfnamefont {S.}~\bibnamefont {{\v{C}}ani{\'c}}},\ }\href@noop {} {\bibfield  {journal} {\bibinfo  {journal} {Bull New Ser Am Math Soc}\ }\textbf {\bibinfo {volume} {58}},\ \bibinfo {pages} {79} (\bibinfo {year} {2021})}\BibitemShut {NoStop}%
\bibitem [{\citenamefont {Man}\ and\ \citenamefont {Doi}(2016)}]{man2016ring}%
  \BibitemOpen
  \bibfield  {author} {\bibinfo {author} {\bibfnamefont {X.}~\bibnamefont {Man}}\ and\ \bibinfo {author} {\bibfnamefont {M.}~\bibnamefont {Doi}},\ }\href@noop {} {\bibfield  {journal} {\bibinfo  {journal} {Phys. Rev. Lett.}\ }\textbf {\bibinfo {volume} {116}},\ \bibinfo {pages} {066101} (\bibinfo {year} {2016})}\BibitemShut {NoStop}%
\bibitem [{\citenamefont {Man}\ and\ \citenamefont {Doi}(2017)}]{man2017vapor}%
  \BibitemOpen
  \bibfield  {author} {\bibinfo {author} {\bibfnamefont {X.}~\bibnamefont {Man}}\ and\ \bibinfo {author} {\bibfnamefont {M.}~\bibnamefont {Doi}},\ }\href@noop {} {\bibfield  {journal} {\bibinfo  {journal} {Phys. Rev. Lett.}\ }\textbf {\bibinfo {volume} {119}},\ \bibinfo {pages} {044502} (\bibinfo {year} {2017})}\BibitemShut {NoStop}%
\bibitem [{\citenamefont {Tabassian}\ \emph {et~al.}(2016)\citenamefont {Tabassian}, \citenamefont {Oh}, \citenamefont {Kim}, \citenamefont {Kim}, \citenamefont {Ryu}, \citenamefont {Cho}, \citenamefont {Koratkar},\ and\ \citenamefont {Oh}}]{tabassian2016graphene}%
  \BibitemOpen
  \bibfield  {author} {\bibinfo {author} {\bibfnamefont {R.}~\bibnamefont {Tabassian}}, \bibinfo {author} {\bibfnamefont {J.-H.}\ \bibnamefont {Oh}}, \bibinfo {author} {\bibfnamefont {S.}~\bibnamefont {Kim}}, \bibinfo {author} {\bibfnamefont {D.}~\bibnamefont {Kim}}, \bibinfo {author} {\bibfnamefont {S.}~\bibnamefont {Ryu}}, \bibinfo {author} {\bibfnamefont {S.-M.}\ \bibnamefont {Cho}}, \bibinfo {author} {\bibfnamefont {N.}~\bibnamefont {Koratkar}},\ and\ \bibinfo {author} {\bibfnamefont {I.-K.}\ \bibnamefont {Oh}},\ }\href@noop {} {\bibfield  {journal} {\bibinfo  {journal} {Nat. Commun.}\ }\textbf {\bibinfo {volume} {7}},\ \bibinfo {pages} {13345} (\bibinfo {year} {2016})}\BibitemShut {NoStop}%
\bibitem [{\citenamefont {Xiao}\ and\ \citenamefont {Wu}(2022)}]{xiao2022droplet}%
  \BibitemOpen
  \bibfield  {author} {\bibinfo {author} {\bibfnamefont {K.}~\bibnamefont {Xiao}}\ and\ \bibinfo {author} {\bibfnamefont {C.-X.}\ \bibnamefont {Wu}},\ }\href@noop {} {\bibfield  {journal} {\bibinfo  {journal} {Phys. Rev. E}\ }\textbf {\bibinfo {volume} {105}},\ \bibinfo {pages} {064609} (\bibinfo {year} {2022})}\BibitemShut {NoStop}%
\bibitem [{\citenamefont {Beroz}\ \emph {et~al.}(2019)\citenamefont {Beroz}, \citenamefont {Hart},\ and\ \citenamefont {Bush}}]{beroz2019stability}%
  \BibitemOpen
  \bibfield  {author} {\bibinfo {author} {\bibfnamefont {J.}~\bibnamefont {Beroz}}, \bibinfo {author} {\bibfnamefont {A.}~\bibnamefont {Hart}},\ and\ \bibinfo {author} {\bibfnamefont {J.~W.}\ \bibnamefont {Bush}},\ }\href@noop {} {\bibfield  {journal} {\bibinfo  {journal} {Phys. Rev. Lett.}\ }\textbf {\bibinfo {volume} {122}},\ \bibinfo {pages} {244501} (\bibinfo {year} {2019})}\BibitemShut {NoStop}%
\bibitem [{\citenamefont {Kingma}\ and\ \citenamefont {Ba}(2014)}]{Kingma2014AdamAM}%
  \BibitemOpen
  \bibfield  {author} {\bibinfo {author} {\bibfnamefont {D.~P.}\ \bibnamefont {Kingma}}\ and\ \bibinfo {author} {\bibfnamefont {J.}~\bibnamefont {Ba}},\ }\href@noop {} {\bibfield  {journal} {\bibinfo  {journal} {CoRR}\ }\textbf {\bibinfo {volume} {abs/1412.6980}} (\bibinfo {year} {2014})}\BibitemShut {NoStop}%
\bibitem [{\citenamefont {LANDAU}\ and\ \citenamefont {LIFSHITZ}(1984)}]{LANDAU198434}%
  \BibitemOpen
  \bibfield  {author} {\bibinfo {author} {\bibfnamefont {L.}~\bibnamefont {LANDAU}}\ and\ \bibinfo {author} {\bibfnamefont {E.}~\bibnamefont {LIFSHITZ}},\ }in\ \href {https://doi.org/https://doi.org/10.1016/B978-0-08-030275-1.50008-4} {\emph {\bibinfo {booktitle} {Electrodynamics of Continuous Media}}},\ Vol.~\bibinfo {volume} {8}\ (\bibinfo {year} {1984})\ pp.\ \bibinfo {pages} {34--85}\BibitemShut {NoStop}%
\bibitem [{\citenamefont {Cheng}\ and\ \citenamefont {Chaddock}(1984)}]{cheng1984deformation}%
  \BibitemOpen
  \bibfield  {author} {\bibinfo {author} {\bibfnamefont {K.}~\bibnamefont {Cheng}}\ and\ \bibinfo {author} {\bibfnamefont {J.}~\bibnamefont {Chaddock}},\ }\href@noop {} {\bibfield  {journal} {\bibinfo  {journal} {Phys. Lett. A}\ }\textbf {\bibinfo {volume} {106}},\ \bibinfo {pages} {51} (\bibinfo {year} {1984})}\BibitemShut {NoStop}%
\bibitem [{\citenamefont {Oh}\ \emph {et~al.}(2008)\citenamefont {Oh}, \citenamefont {Ko},\ and\ \citenamefont {Kang}}]{oh2008shape}%
  \BibitemOpen
  \bibfield  {author} {\bibinfo {author} {\bibfnamefont {J.~M.}\ \bibnamefont {Oh}}, \bibinfo {author} {\bibfnamefont {S.~H.}\ \bibnamefont {Ko}},\ and\ \bibinfo {author} {\bibfnamefont {K.~H.}\ \bibnamefont {Kang}},\ }\href@noop {} {\bibfield  {journal} {\bibinfo  {journal} {Langmuir}\ }\textbf {\bibinfo {volume} {24}},\ \bibinfo {pages} {8379} (\bibinfo {year} {2008})}\BibitemShut {NoStop}%
\bibitem [{\citenamefont {Oh}\ \emph {et~al.}(2010)\citenamefont {Oh}, \citenamefont {Ko},\ and\ \citenamefont {Kang}}]{oh2010analysis}%
  \BibitemOpen
  \bibfield  {author} {\bibinfo {author} {\bibfnamefont {J.~M.}\ \bibnamefont {Oh}}, \bibinfo {author} {\bibfnamefont {S.~H.}\ \bibnamefont {Ko}},\ and\ \bibinfo {author} {\bibfnamefont {K.~H.}\ \bibnamefont {Kang}},\ }\href@noop {} {\bibfield  {journal} {\bibinfo  {journal} {Phys. Fluids}\ }\textbf {\bibinfo {volume} {22}},\ \bibinfo {pages} {032002} (\bibinfo {year} {2010})}\BibitemShut {NoStop}%
\end{thebibliography}
\end{document}